\documentstyle[12pt]{article}
\begin{document}
\pagestyle{plain}
\setcounter{page}{1}
\newcounter{bean}
\baselineskip16pt
\begin{titlepage}

\begin{flushright}
PUPT-1644\\
hep-th/9608108
\end{flushright}
\vspace{20 mm}

\begin{center}
{\huge Emission of charged particles from }

\vspace{5mm}
{\huge four- and five-dimensional black holes}

\end{center}

\vspace{10 mm}

\begin{center}
{\large Steven S.~Gubser\footnote{e-mail: {\tt ssgubser@puhep1.princeton.edu}} 
and Igor R.~Klebanov\footnote{e-mail: {\tt klebanov@puhep1.princeton.edu}}
}

\vspace{3mm}

Joseph Henry Laboratories\\
Princeton University\\
Princeton, New Jersey 08544

\end{center}

\vspace{2cm}

\begin{center}
{\large Abstract}
\end{center}

\noindent
Recently Das and Mathur found that the leading order Hawking emission
rate of neutral scalars by near-extremal $D=5$ black holes is exactly
reproduced by a string theoretic model involving intersecting D-branes.
We show that the agreement continues to hold for charged scalar
emission.  We further show that similar agreement can be obtained for 
a class of near-extremal $D=4$ black holes using a model inspired by 
M-theory.  In this model, BPS saturated $D=4$ black holes with four charges 
are realized in M-theory as 5-branes triply intersecting over a string.
The low-energy excitations are signals traveling on the intersection
string.

\vspace{2cm}
\begin{flushleft}
August 1996
\end{flushleft}
\end{titlepage}
\newpage
\renewcommand{\baselinestretch}{1.1} 

\renewcommand{\epsilon}{\varepsilon}
\def\fixit#1{}
\def\comment#1{}
\def\equno#1{(\ref{#1})}
\def\equnos#1{(#1)}
\def\sectno#1{section~\ref{#1}}
\def\figno#1{Fig.~(\ref{#1})}
\def\D#1#2{{\partial #1 \over \partial #2}}
\def\df#1#2{{\displaystyle{#1 \over #2}}}
\def\tf#1#2{{\textstyle{#1 \over #2}}}
\def\d{{\rm d}}
\def\e{{\rm e}}
\def\i{{\rm i}}
\def\Leff{L_{\rm eff}}

\section{Introduction}
\label{Intro}

During recent months, impressive progress has been made 
towards a more fundamental explanation of the semi-classical properties
of black holes. Strominger and Vafa found a statistical explanation of
the Bekenstein-Hawking entropy \cite{jb,swh}
for a class of supersymmetric black
holes in $D=5$ \cite{sv}.  When such black holes carry three different
$U(1)$ charges, their horizon area is non-vanishing even for the
extremal solution.  In string theory, the Bekenstein-Hawking entropy
of macroscopic black holes is exactly reproduced by the counting of
states of a configuration of Dirichlet branes \cite{polchnotes}
which carries the same charges \cite{sv}.  This result was extended to
$D=5$ black holes near extremality in \cite{cm,hs1}. Furthermore, in
\cite{cm} it was shown how Hawking emission from the near-extremal
black holes takes place in the stringy description.  The model
involves $n_1$ 1-branes marginally bound to $n_5$ 5-branes, with some
longitudinal momentum along the 1-branes carried by left moving
open strings. Near extremality, a small number of right movers is also
present, so that a left moving and a right moving open string may
collide to produce an outgoing closed string \cite{cm,hk}. In
\cite{cm} it was shown that this mechanism leads to a thermal
distribution for the massless outgoing particles, as expected of 
Hawking radiation. The inverse of this process, which gives the
leading order contribution to the absorption of closed strings, was
also found to be in agreement with the semi-classical gravity, up to
an overall normalization \cite{dmw}.  More recently, in an impressive
paper \cite{dm} Das and Mathur carefully normalized the leading
emission and absorption rates, both in semi-classical gravity and
in the D-brane picture, and found perfect agreement!  The specific picture
used in \cite{dm} follows that suggested in \cite{dm1,ms}:  the low-energy
dynamics of the D-brane configuration is captured by a single string with
winding number $n_1 n_5$ which is free to vibrate only within the 5-brane
hyperplane.  Vibrations in the transverse directions are not allowed
because the 1-branes are bound to the 5-branes, so only transverse scalars
are emitted at low energies, in agreement with semi-classical gravity.

In this paper we extend the result of \cite{dm} in two directions.
First we generalize the calculation to outgoing scalars which are
charged and massive (their mass is proportional to charge according to
the BPS condition).  The $D=5$ black hole carries three different
charges which in the string context are realized as the number of
1-branes, the number of 5-branes, and the Kaluza-Klein charge
(the momentum along the 1-brane direction).  Endowing the
outgoing scalars with the Kaluza-Klein charge in both the D-brane and
the gravity calculations, we once again find perfect agreement.

Our second extension is to supersymmetric four-dimensional black holes
with regular horizons. In \cite{cy,ct} it was shown that for an
extremal $D=4$ black hole to have a finite horizon area, it must carry
four different $U(1)$ charges.  Such black holes can be embedded into
string theory, but the necessary configurations involve solitonic
5-branes or Kaluza-Klein monopoles in addition to the
D-branes \cite{sm,jkm}.  In \cite{at,kt,gkt} it was argued that it is
advantageous to view the $D=4$ black holes as dimensionally reduced
configurations of intersecting branes in M-theory.  A specific
configuration useful for explaining the Bekenstein-Hawking entropy is
the $5\bot 5\bot 5$ intersection \cite{kt}: there are $n_1$ 5-branes
in the $(12345)$ hyperplane, $n_2$ 5-branes in the $(12367)$
hyperplane, and $n_3$ 5-branes in the $(14567)$ hyperplane. One also
introduces a left moving momentum along the intersection
string (in the $\hat 1$ direction). If the length of this direction is
$L_1$, then the momentum is quantized as $2\pi n_K/L_1$, 
so that $n_K$ plays the role of the fourth $U(1)$
charge.  Upon compactification on $T^7$ the metric of the $5\bot 5\bot
5$ configuration
reduces to that of the $D=4$ black hole with four charges.  Just like
in the D-brane description of the $D=5$ black hole, the low-energy
excitations are signals propagating along the intersection string. In
M-theory the relevant states are likely to be small 2-branes with
three holes glued into the three different hyperplanes \cite{kt}. As a
result, the effective winding number of the intersection string is
$n_1 n_2 n_3$. This fact, together with the assumption that these
modes carry central charge $c=6$, is enough to reproduce the
Bekenstein-Hawking entropy, $S=2\pi\sqrt{n_1 n_2 n_3 n_K}$ \cite{kt}.
In this paper we go further and show that this ``multiply-wound
string'' model of the four-charge $D=4$ black hole is also capable of 
accounting for Hawking radiation.  It exactly reproduces the emission rate
found using the methods of semi-classical gravity for scalars carrying 
Kaluza-Klein charge.

\section{The string theory analysis}
\label{String}

Let us start by recapitulating in a streamlined way the string theory
treatment of the five-dimensional black hole \cite{dm,dmw,ms}. 
The essential assumption is that the $n_1$ D-strings
bound to $n_5$ 5-branes act as a single D-string of winding number
$n_1 n_5$, which is free to move only within the 5-brane
hyperplane. This multiply-wound D-string is described by the following
six-dimensional effective action:

\begin{eqnarray}
S &=& T_{\rm D} 
      \int \d^2 \xi \, \e^{-\phi} \sqrt{\det \left[ G_{\alpha\beta}(X) +
        B_{\alpha\beta}(X) + F_{\alpha\beta} \right]} +
      \df{1}{2 \kappa_6^2} \int \d^6 x \, \sqrt{-g}
         \e^{-2 \phi} R + \ldots  \nonumber \\
  &=& T_{\rm D}
      \int \d^2 \xi \, \tf{1}{2} (\delta_{ij} + 2 \kappa_6 h_{ij})
        \partial_\alpha X^i \partial^\alpha X^j +
      \int \d^6 x \, \tf{1}{2} (\partial_M h_{ij}) (\partial^M h^{ij}) +
      \ldots \ .                                     \label{StringAction}
\end{eqnarray}

\noindent 
Here $\alpha$ and $\beta$ are coordinates on the D-string worldsheet,
$i$ and $j$ run over the coordinates $y_2,\ldots y_5$ compactified on
$T^4$, and $M$ and $N$ run over the six other coordinates. The spatial
coordinate $y_1$ is parallel to the D-string, whose tension we denote
by $T_D$.  In the first line of \equno{StringAction}, the D-string
action is of the standard Dirac-Born-Infeld form, while the gravity
action is part of the standard type~IIB action.

In \equno{StringAction} we have suppressed those parts of the action
not directly relevant to the calculations below: in the first line,
kinetic terms for the dilaton and the antisymmetric tensor are
omitted, as well as the fermion terms both on the D-string and in the
bulk of spacetime; in the second line, the only terms listed are the
leading order interaction term coupling the bosonic excitations of the
D-string with the internal gravitons $h_{ij}$ (scalars in six
dimensions) and the kinetic terms for these fields.

One more subtlety involves the range of the coordinate $y_1$.
Strictly speaking, the action in \equno{StringAction} describes only
the vibrations of a single 1-brane within the world-volume of a single
5-brane.  When we compactify down to five dimensions by wrapping the
D-string around a circle of circumference $L_1$, $y_1$ becomes
periodic in the obvious manner, $y_1 \sim y_1 + L_1$.  There is a neat
trick \cite{dm,dm1} for handling the case of multiple 1-{} and
5-branes: the effective theory in five dimensions is obtained simply
by identifying $y_1 \sim y_1 + L_{\rm eff}$ where $L_{\rm eff} = n_1
n_5 L_1$.  This prescription corresponds to the entropically favored
configuration of a single 1-brane and a single 5-brane wound $n_1$ and
$n_5$ times, respectively \cite{ms}.
Note that the total momentum of the D-string is quantized in units of
$2 \pi / L_1$, as is the momentum parallel to the D-string of
particles in the bulk of spacetime.  By contrast, the bosonic excitations 
of the D-string, which are described by quanta of the
fields $X^i$, have momentum quantized in units of $2 \pi / \Leff$.

The numbers of left moving ($E=-p$) and right moving ($E=p$) bosonic
excitations are assumed to follow separate thermal distributions

\begin{equation}
\rho_L(p_0) = \df{1}{\e^{\beta_L p_0} - 1} \qquad
\rho_R(q_0) = \df{1}{\e^{\beta_R q_0} - 1}               \label{RhoLR}
\end{equation}

\noindent
with $T_R \ll T_L$.  Equivalently, \equno{RhoLR} can be thought of as
a single thermal distribution for all the modes subject to a
constraint on the total momentum imposed via a chemical potential, as
explained for instance in \cite{dm}.

We have now finished setting the stage for the analysis of the
dominant decay processes of the D-string.  It is remarkable that the
relatively complex string theoretic structure of multiply wound
intersecting 1-{} and 5-branes with a condensate of open strings
running along the intersection should boil down to such a simple
description as a single long string described by the action on the
second line of \equno{StringAction}.  While this picture undoubtedly
needs a more precise justification, it works very well. It correctly
predicts the black hole entropy \cite{ms}, the neutral particle
emission \cite{dm}, and, as we will show, the charged particle
emission.

The invariant amplitude for the process where a left moving $X^i$
excitation with momentum $p$ and a right moving $X^j$ excitation with
momentum $q$ collide and turn into a graviton with polarization in the $ij$
direction and momentum $k$ is 

\begin{equation}
{\cal M} = \sqrt{2} \kappa_6 p \cdot q \ .           \label{Amp5d}
\end{equation}

\noindent 
Let us fix the graviton's momentum parallel to the D-string:
$k_1 = -e$ where $e$ is the Kaluza-Klein charge in five dimensions.  (We
conventionally take $e>0$ for left movers in order that when $T_R \ll T_L$
the black hole has an overall positive charge.)
The total rate to produce such a
graviton with definite transverse momentum $\vec{k}$ follows directly 
from \equno{Amp5d} and kinematical arguments:

\begin{eqnarray}
\Gamma (\vec{k}) \df{\d^4 k}{(2 \pi)^4} &=&
  2 \int_{-\infty}^0 \df{\Leff \d p_1}{2 \pi} \, \rho_L(-p_1)
    \int_0^\infty \df{\Leff \d q_1}{2 \pi} \, \rho_R(q_1)  \nonumber \\
 & & \qquad\ \cdot (2 \pi)^2 \delta (p_0 + q_0 - k_0) \delta (p_1 + q_1 - k_1)
 {1\over 8 L_1 \Leff}
   \df{|{\cal M}|^2}{p_0 q_0 k_0} \df{\d^4 k}{(2 \pi)^4}  \nonumber \\
 &=& \df{\kappa_5^2 \Leff}{4} \df{k^2}{\omega}
  \rho_L \left( \df{\omega + e}{2} \right)
  \rho_R \left( \df{\omega - e}{2} \right)
  \df{\d^4 k}{(2 \pi)^4}                             \label{GammaUniv}
\end{eqnarray}

\noindent 
In this equation and in the following, $k = |\vec{k}|$ is
the magnitude of the particle's momentum in the four noncompact
spatial dimensions.  The factor of $2$ outside the integral in the
first line of \equno{GammaUniv} accounts for the fact that the same
graviton can be produced by a left moving $X^i$ quantum colliding with
a right moving $X^j$ quantum or by a left moving $X^j$ quantum
colliding with a right moving $X^i$ quantum.  To obtain a rate which
can be directly compared with semi-classical calculations, we make the
crucial high temperature expansion

\begin{equation}
\rho_L(E) \approx {1\over \beta_L E }                   \label{RhoLExpand}
\end{equation}

\noindent
and use the results of previous papers \cite{sv,cm,dm,ms} to express the 
right and left moving temperatures in terms of properties of the classical
five-dimensional near-extremal geometry:

\begin{equation}
T_L = \df{S_L}{\pi \Leff} \approx \df{S_{\rm BH}}{\pi \Leff} = 
  \df{A_{\rm h}}{4 \pi \Leff G_5}\ ,\qquad
T_R = \tf{1}{2} T_{\rm H} \ .
\label{temps}\end{equation}

\noindent
In normalizing the left moving temperature we used the fact that there 
are four species of massless bosons and four species of massless fermions
on the string (i.e.{} the central charge is $c=6$). 
Now the differential rate of emission of a given species of scalar particle
of charge $e$ and mass $m=|e|$ is

\begin{equation}
\Gamma(\omega) \df{\d \omega}{2 \pi} = \df{1}{8 \pi^2} 
  \df{A_{\rm h} k^2 (\omega - e)}{\e^{\beta_{\rm H} (\omega - e)} - 1}
\d \omega \ .
                                                   \label{Gamma5d}
\end{equation}

\noindent
Note that the ten possible polarizations of $h_{ij}$ give ten different
species of scalars.  For a given species, the total power radiated for 
frequencies between $\omega_1$ and $\omega_2$ is

\begin{equation}
P(\omega_1,\omega_2) = \int_{\omega_1}^{\omega_2} 
  \omega \Gamma(\omega) \df{\d \omega}{2 \pi} \ .
\end{equation}

In the next section we carry out a parallel calculation of the charged
scalar emission rate by solving the scalar field equation in the classical
black hole background.

\section{The five-dimensional case}
\label{FiveDim}

The main result of \cite{dm} was to show that for $e = 0$,
\equno{Gamma5d} can be reproduced by a semi-classical calculation in 
the long wavelength limit $Q_i k^2 \ll 1$.  Our goal in this section 
is to extend this result to the case of charged 
particles.\footnote{We thank J.~Maldacena and  
A.~Strominger for informing us of their
independent verification of the charged particle case \cite{mprivate}.}

Let us start by reviewing the classical geometry \cite{at1,cm,hs1}.  
The 1-{} and 5-brane configuration in ten dimensions is described by
the following string metric and dilaton:

\begin{eqnarray}
\d s_{(10,{\rm str})}^2 &=& \df{1}{\sqrt{f_1 f_5}} 
  \left( -\d t^2 + \d y_1^2 + K (\d t + \d y_1)^2 \right) + 
  \sqrt{f_1/f_5} \left( \d y_2^2 + \ldots + \d y_5^2 \right) +  \nonumber \\
 & &  \sqrt{f_1 f_5} \left( \d r^2 + r^2 \d \Omega_{S^3}^2 \right)  
         \nonumber \\
\e^{-2 \phi_{(10)}} &=& f_5/f_1  \ ,
\end{eqnarray}

\noindent
where

\begin{equation}
f_1 = 1 + {Q_1\over r^2}
\ , \qquad f_5 = 1 + {Q_5\over r^2}\ , \qquad K = {Q_K\over r^2}\ .
\end{equation}

\noindent
The charges $Q_1$ and $Q_5$ are proportional to the numbers of 1-{} and 
5-branes. In the following, we will only need the normalization of the product
of the two charges \cite{cm,kt}:

\begin{equation}
Q_1 Q_5 = {\kappa_5^2 L_1 n_1 n_5\over 4 \pi^3}
\ ,\end{equation}

\noindent
where $\kappa_5^2=\kappa_{10}^2/\prod_{i=1}^5 L_i$.
The Kaluza-Klein charge is also quantized \cite{cm,kt}:

\begin{equation}
Q_K = {\kappa_5^2 n_K \over \pi L_1} \ .
\end{equation}

Reducing to six dimensions by compactifying $y_2, \ldots, y_5$ on a 
four-torus $T^4$ yields the metric which describes a string pointing
along the $\hat 1$-direction:

\begin{equation}
\d s_{(6)}^2 = \df{1}{\sqrt{f_1 f_5}} 
   \left( -\d t^2 + \d y_1^2 + K (\d t + \d y_1)^2 \right) +
  \sqrt{f_1 f_5} \left( \d r^2 + r^2 \d \Omega_{S^3}^2 \right) \ .  
                                                        \label{6DimMetric}
\end{equation} 

\noindent
The six-dimensional dilaton is constant.  Reduction from the string in
$D=6$ to the black hole in $D=5$ is achieved by comparing
\equno{6DimMetric} to the form

\begin{equation}
\d s_{(6)}^2 = \e^{-2D/3} \d s_{(5)}^2 +
  \e^{2D} \left( \d y_1 + A_\mu \d x^\mu \right)^2
\ , \label{KKRed} \end{equation}

\noindent
where the $x^\mu$ are coordinates for the five noncompact directions.  The 
factors of $\e^D$ in \equno{KKRed} are arranged to put the 
five-dimensional metric in Einstein frame.  The result is

\begin{eqnarray}
\d s_{(5)}^2 &=& -f^{-2/3} \d t^2 +
                  f^{1/3} \left( \d r^2 + r^2 \d \Omega_{S^3}^2 \right)
\nonumber \\
f &=& \left( 1 + {Q_1\over r^2} \right)
      \left( 1 + {Q_5\over r^2} \right)
      \left( 1 + {Q_K\over r^2} \right)  \nonumber \\
A_0 &=& K/(1+K) = Q_K/(Q_K+r^2) \ .                         \label{5DMetric}
\end{eqnarray}

\noindent
We have chosen the normalization of $A_0$ so that $m = |e|$ 
for BPS saturated charged scalars, as in \sectno{String}.
The horizon area and Bekenstein-Hawking entropy following from 
\equno{5DMetric} are 

\begin{eqnarray}
A_{\rm h} &=& 2 \pi^2 \prod_{i=1}^3 \sqrt{Q_i}  \nonumber \\
S_{\rm BH} &=& {A_{\rm h} \over 4 G_5} = 2 \pi \prod_{i=1}^3 \sqrt{n_i} \ ,
                                                         \label{Area5d}
\end{eqnarray}

\noindent
where for convenience we have set $Q_2 = Q_5$, $Q_3 = Q_K$,
$n_2 = n_5$, and $n_3 = n_K$. 

To describe slight departures from extremality, one introduces a small
parameter $\mu$ with the same dimensions ($[{\rm length}]^2$) as the 
$Q_i$ \cite{kt1,ct1}.  The near-extremal entropy has the form
characteristic of $1+1$ dimensional field theory if we choose the charges
in the following way:

\begin{equation}
\mu \ll Q_K \ll Q_1, Q_5 \ .
\end{equation}

\noindent
This condition is necessary to insure that the anti-onebranes and 
anti-fivebranes \cite{hms} are suppressed, so that the departure 
from extremality is due only to the
right movers on the intersection string. The changes in the entropy
and the Hawking temperature are then given by \cite{dm}

\begin{equation}
\Delta S= \sqrt{\pi E \Leff}\ , \qquad T_{\rm H} = 2 \sqrt{E\over \pi \Leff}
\ , \label{1Dent}\end{equation}

\noindent
where $E= M- M_0$.

When the five-dimensional black hole is raised slightly above extremality,
the dominant Hawking radiation processes are those where scalars are emitted 
in an $s$-wave.  The $s$-wave is dominant because the near-extremal black hole
emits mostly particles with wavelength much longer than its typical length
scales $\sqrt{Q_i}$.  Emission rates for higher partial waves and for higher
spin particles are expected to be suppressed by powers of $A_{\rm h} /
\lambda^3$.  The analysis of \sectno{String} enables us to compare to string
theory the emission rate of scalars coming from the $T^4$ polarizations of 
the ten-dimensional graviton.

The Hawking rate is determined by the classical absorption probability, 
which for the case at hand may be calculated using the extremal 
geometry~\cite{dm}.  Intuitively speaking, the reason why the extremal
geometry may be used is because its horizon has finite area and finite
electrostatic potential which are corrected only at order $\mu$ in the 
near extremal case: consequently the classical absorption probability also 
suffers only $O(\mu)$ departures from its value at extremality.

The classical field equation in five dimensions for an $s$-wave scalar
follows from plugging the ansatz $\phi(t,y_1,r) = \e^{-\i \omega t} 
\e^{-\i e y_1} R(r)$ into the six-dimensional Laplace equation:

\begin{equation}
\Box_{(6)} \phi = \df{1}{\sqrt{-g^{(6)}}} \partial_M \sqrt{-g^{(6)}}
  g_{(6)}^{MN} \partial_N\phi = 0 \ .
\end{equation}

\noindent
The radial equation takes the remarkably simple form 

\begin{equation}
\left[ (\omega - e A_0)^2 - \df{m^2}{(1+K)^2} + \df{1}{f r^3}
  \df{\d}{\d r} r^3 \df{\d}{\d r} \right] R(r) = 0 \ .       \label{REq5d}
\end{equation}

\noindent
Following the method developed by Unruh \cite{unruh} and used in 
\cite{dmw,dm}, we solve the radial equation to leading order in the small
quantities $Q_i k^2$ and extract the classical absorption probability.  
The solution is achieved by patching together three regions:

\setcounter{bean}{0}
\begin{list}{{\bf \Roman{bean}. }}{\usecounter{bean}}
\item For $r \ll \sqrt{Q_i}$, that is to say very close to the horizon, 
the dominant terms of \equno{REq5d} are

\begin{equation}
\left( \df{1}{r^3} \df{\d}{\d r} r^3 \df{\d}{\d r} + \df{P}{r^6} \right)
  R(r) = 0                                              \label{NearEq5d}
\end{equation}

\noindent
where

\begin{equation}
P = (\omega - e a_0)^2 \prod_{i=1}^3 Q_i 
\end{equation}

\noindent
and $a_0 = A_0\big|_{r=0} = 1$ is the electrostatic potential on the horizon.
The solution which determines classical absorption by the black hole must
represent purely infalling matter close to the horizon.  The infalling 
solution to \equno{NearEq5d} is

\begin{equation}
R(r) = \e^{\i \sqrt{P} / (2 r^2)} \ .               \label{ISol5d}
\end{equation}

\item The key to matching the near and far regions is to use the 
long wavelength limit to trivialize \equno{REq5d} for $r \sim \sqrt{Q_i}$.
The result is

\begin{equation}
\df{1}{r^3} \df{\d}{\d r} r^3 \df{\d}{\d r} R(r) = 0 \ ,
\end{equation}

\noindent
and the solution is of the form

\begin{equation}
R(r) = C + D/r^2 \ .                                   \label{IISol5d}
\end{equation}

\item For $r \gg \sqrt{Q_i}$, an expansion in powers of $1/r$ up to 
$1/r^2$ yields 

\begin{equation}
\left [ \df{1}{r^3} \df{\d}{\d r} r^3 \df{\d}{\d r} + 
  k^2 \left( 1 + \df{\sum_{i=1}^3 Q_i - 2 e Q_3 / (\omega + e)}{r^2} \right) 
 \right ] R(r) = 0
\end{equation}

\noindent
where $k^2 = \omega^2 - m^2$.  The general solution is 

\begin{eqnarray}
R(r) &=& \df{\alpha J_{\nu}(kr) + \beta J_{-\nu}(kr) }{ kr}  \nonumber \\
\nu &=& \sqrt{1 - k^2 
\left( \sum_{i=1}^3 Q_i - 2 e Q_3 / (\omega + e) \right)} \ .
                                                      \label{IIISol5d}
\end{eqnarray}
\end{list}

\noindent
To obtain matching between ${\bf II}$ and ${\bf III}$ it is necessary to
have $\nu$ close to $1$.  For $e>0$, $\nu = 1 - O(Q_i k^2)$, so matching is
insured by the long wavelength limit.  For $e<0$, $1-\nu$ is not small 
unless also $Q_K m \omega \ll 1$.  The condition $Q_K m \omega \ll 1$ is
necessary to insure that the particle is perturbatively scattered:  if it 
fails, then resonance with bound states affects scattering processes 
significantly, and re-absorption must be taken into account in Hawking 
emission calculations.

Matching the three regions to leading order in $1-\nu$, we find

\begin{equation}
\alpha = 2 \qquad \beta = \df{\i k^2 \sqrt{P}}{4 (1-\nu)} \ .
\end{equation}

\noindent
We have allowed the normalization of $R(r)$ to be fixed by the coefficient
on the near-horizon solution \equno{ISol5d}.  This normalization is
arbitrary, but for the purpose of determining the S-matrix element, only
the relative coefficient between outgoing and ingoing waves far from the
black hole is relevant.  This relative coefficient is determined by the
ratio $\beta/\alpha$, as we shall see.

The standard asymptotic form for an $s$-wave that one derives from a 
partial wave expansion is

\begin{equation}
R(r) \sim \df{S_0 \e^{\i kr} - \i \e^{-\i kr}}{(kr)^{3/2}}  \qquad 
 {\rm as} \ r \to \infty \ .                          \label{SWaveStandard}
\end{equation}

\noindent
Comparison of \equno{SWaveStandard} with the $r \to \infty$ asymptotics 
of \equno{IIISol5d} allows one to read off the S-matrix element:

\begin{equation}
S_0 = \e^{\i \pi (1-\nu)} 
       \df{1 -\tf{\beta}{\alpha} \e^{-\i \pi (1-\nu)}  }{
           1 -\tf{\beta}{\alpha} \e^{\i \pi (1-\nu)}  } \ .
\end{equation}

\noindent
The classical absorption probability is 

\begin{equation}
1-|S_0|^2 \approx \df{\pi}{2} k^2 (\omega - e a_0) \prod_{i=1}^3 \sqrt{Q_i} 
  = \df{k^2 (\omega - e a_0) A_{\rm h}}{4 \pi} \ .
\end{equation}

\noindent
The Hawking rate is computed from the classical absorption probability 
via a standard formula of quantum field theory in curved spacetime 
\cite{gibbons}:

\begin{equation}
\Gamma(\omega) {\d \omega\over 2\pi} = \df{1 - |S_0|^2 }{ 
   \e^{\beta_{\rm H} (\omega - e a_0)} - 1} \df{\d \omega}{2 \pi} 
 \approx \df{1}{8 \pi^2} \df{k^2 (\omega - e a_0) A_{\rm h} }{ 
   \e^{\beta_{\rm H} (\omega - e a_0)} - 1} \d \omega \ ,    \label{Hawk5d}
\end{equation}

\noindent
in exact agreement with \equno{Gamma5d}
since $a_0=1$ at extremality. 
Note how the electrostatic 
potential on the horizon, $a_0$, enters 
as a chemical potential.
We have verified that \equno{Hawk5d}
holds far from extremality as well, provided $\mu m^2 \ll 1$. 
For the near-extremal geometry, $a_0=1-O(\mu)$ where $\mu$
is the parameter measuring deviation from extremality.
It would be interesting to see how the D-brane picture
reproduces the $O(\mu)$ corrections.

It is worth pointing out two simple physical properties of \equno{Hawk5d}.
First, Hawking
emission of particles with $e>0$ is enhanced relative to $e<0$ particles
of the same mass, whereas the classical absorption of the former is 
suppressed relative to the latter.  This is just what one expects from a
black hole with positive charge and hence $a_0>0$.  Second, the phenomenon
of super-radiance cannot occur here because $\omega - e a_0 \geq \omega - m
\geq 0$:  in effect it is forbidden by supersymmetry.

\section{The four-dimensional case}
\label{FourDim}

The semi-classical computation for the four-dimensional case is another
straightforward application of the methods developed in \cite{unruh}.
As in the five-dimensional case, it suffices to compute the S-matrix
element for long wavelength $s$-wave scattering to zeroth order in
$\mu$, that is to say for the extremal black hole.  The 11-dimensional
configuration of three sets of 5-branes intersecting along a common
1-brane is described by the metric \cite{kt1}

\begin{eqnarray}
\d s_{(11)}^2 &=& (f_1 f_2 f_3)^{-1/3}
  \left[ -\d t^2 + \d y_1^2 + K (\d t+ \d y_1)^2
\right] +  \nonumber \\
 & & \ \ (f_1 f_2 f_3)^{-1/3}
  \left[ f_3 \left( \d y_2^2 + \d y_3^2 \right) + 
         f_2 \left( \d y_4^2 + \d y_5^2 \right) + 
         f_1 \left( \d y_6^2 + \d y_7^2 \right) \right] +  \nonumber \\
 & & \ \ (f_1 f_2 f_3)^{2/3}
  \left( \d r^2 + r^2 \d \Omega_{S^2}^2 \right)  \nonumber
\end{eqnarray}
\begin{equation}
f_i = 1 + \df{Q_i}{r} \ , \qquad K = \df{Q_K}{r} \ .     \label{11DimMetric}
\end{equation}

\noindent 
The charges $Q_i$ are related to the numbers of 5-branes \cite{kt}:

\begin{equation}
Q_1 = \df{n_1}{L_6 L_7} \left( \df{\kappa_{11}}{4 \pi} \right)^{2/3} \qquad 
Q_2 = \df{n_1}{L_4 L_5} \left( \df{\kappa_{11}}{4 \pi} \right)^{2/3} \qquad
Q_3 = \df{n_1}{L_2 L_3} \left( \df{\kappa_{11}}{4 \pi} \right)^{2/3} \ ,
\end{equation}

\noindent
where $L_i$ is the range of the coordinate $y_i$.
The Kaluza-Klein charge $Q_K$ is quantized as
\cite{kt}

\begin{equation}
Q_K =  \kappa^2_4 \df{n_K}{L_1} \ ,                 \label{KKCharge}
\end{equation}

\noindent
where $\kappa_4^2= \kappa_{11}^2/\prod_{i=1}^7 L_i$.

To obtain the string in five dimensions one compactifies on a
six-torus involving the coordinates $y_2, \ldots, y_7$. 
This gives the following metric:

\begin{equation}
\d s_{(5)}^2 = (f_1 f_2 f_3)^{-1/3} 
    \left[ -\d t^2 + \d y^2 + K \left( \d t + \d y \right)^2 \right] + 
   (f_1 f_2 f_3)^{2/3} \left( \d r^2 + r^2 \d \Omega_{S^2}^2 \right)  
                                                    \label{5DimMetric}
\end{equation}

\noindent
while the five-dimensional dilaton turns out to be constant.
Reduction from the string in $D=5$ to the black hole in
$D=4$ is achieved by comparing \equno{5DimMetric} 
to the form

\begin{equation}
\d s_{(5)}^2 = \e^{-D} \d s_{(4)}^2 + 
  \e^{2D} \left( \d y_1 + A_\mu \d x^\mu \right)^2 \ ,   \label{KKReduce}
\end{equation}

\noindent
where the factors of $\e^D$ are arranged to put the four-dimensional metric
in Einstein frame.  The results are quite similar to the five-dimensional
black hole:

\begin{eqnarray}
\d s_{(4)}^2 &=& -f^{-1/2} \d t^2 + 
                  f^{1/2} \left( \d r^2 + r^2 \d \Omega_{S^2}^2 \right)
\nonumber \\
f &=& \prod_{i=1}^4 \left( 1 + {Q_i\over r} \right)  \nonumber \\
A_0 &=& K/(1+K) = Q_K/(Q_K+r)                           \label{4DimMetric}
\end{eqnarray}

\noindent
where we have set $Q_4 = Q_K$.  The horizon area and Bekenstein-Hawking
entropy are given by 

\begin{eqnarray}
A_{\rm h} &=& 4 \pi \prod_{i=1}^4 \sqrt{Q_i}  \nonumber \\
S_{\rm BH} &=& {A_{\rm h} \over 4 G_4} = 2 \pi \prod_{i=1}^4 \sqrt{n_i} \ .
                                                         \label{Area4d}
\end{eqnarray}

\noindent
As shown in \cite{kt1}, the near-extremal entropy has the form
\equno{1Dent} characteristic of $1+1$ dimensional field theory if
the charges satisfy

\begin{equation}
\mu \ll Q_K \ll Q_1, Q_2, Q_3 \ , 
\end{equation}

\noindent
where again $\mu$ parametrizes small departures from extremality.
The Hawking temperature is also given by \equno{1Dent} with
$\Leff= L_1 n_1 n_2 n_3$.

As in the case of the five-dimensional black hole and for the same reasons,
the dominant Hawking radiation processes for this four-dimensional black 
hole are those where a four-dimensional scalar is emitted in an $s$-wave.
The classical field equation
has almost the same form as in the five-dimensional case: plugging the 
ansatz $\phi(t,y_1,r) = \e^{-\i \omega t} \e^{-\i e y_1} R(r)$ into the 
equation

\begin{equation}
\Box_{(5)} \phi = \df{1}{\sqrt{-g^{(5)}}} \partial_M \sqrt{-g^{(5)}} 
  g_{(5)}^{MN} \partial_N \phi = 0 \ ,
\end{equation}

\noindent
one obtains 

\begin{equation}
\left[ (\omega - e A_0)^2 - \df{m^2}{(1+K)^2} + \df{1}{f r^2} 
  \df{\d}{\d r} r^2 \df{\d}{\d r} \right] R(r) = 0 \ .       \label{REq4d}
\end{equation}

\noindent
We have again chosen conventions so that $m = |e|$.
One can solve \equno{REq4d} to leading order in the small quantities 
$Q_i k$ (where $k^2 = \omega^2 - m^2$) using simpler approximations than in
the five-dimensional case.  The three regions can be treated as follows:

\setcounter{bean}{0}
\begin{list}{{\bf \Roman{bean}. }}{\usecounter{bean}}
\item In the near region, the dominant terms of \equno{REq4d} are

\begin{equation}
\left( \df{1}{r^2} \df{\d}{\d r} r^2 \df{\d}{\d r} + \df{P}{r^4} \right) 
  R(r) = 0 
\end{equation}
where

\begin{equation}
P = (\omega - e a_0)^2 \prod_{i=1}^4  Q_i \ .
\end{equation}

\noindent
The infalling solution is 

\begin{equation}
R(r) = \e^{\i \sqrt{P}/r} \ .                           \label{ISol4d}
\end{equation}

\item In the intermediate region, the same long wavelength limit as used in 
the five-dimensional case makes the equation trivial:

\begin{equation}
\df{1}{r^2} \df{\d}{\d r} r^2 \df{\d}{\d r} R(r) = 0 \ .
\end{equation}

\noindent
The solution is of the form

\begin{equation}
R(r) = C + D/r \ .                                   \label{IISol4d}
\end{equation}

\item Far from the black hole, it turns out to be sufficient to use 
the free particle equation, 

\begin{equation}
\left( \df{1}{r^2} \df{\d}{\d r} r^2 \df{\d}{\d r} + k^2 \right) R(r) = 0 \ .
\end{equation}

\noindent
The general solution is

\begin{equation}
R(r) = \alpha \df{\sin kr}{kr} - \beta \df{\cos kr}{kr} \ .
                                                     \label{IIISol4d}
\end{equation}

\end{list}

\noindent
To insure perturbative scattering it is necessary to assume 
$Q_K m \omega / k \ll 1$ when $e<0$.  Matching the three regions yields 

\begin{equation}
\alpha = 1 \qquad 
\beta = -\i k \sqrt{P} \ .
\end{equation}

\noindent
Comparing the solution \equno{IIISol4d} with the standard asymptotic form

\begin{equation}
R(r) \sim \df{S_0 \e^{\i kr} - \e^{-\i kr}}{kr}  \qquad 
 {\rm as} \ r \to \infty
\end{equation}

\noindent
yields for the S-matrix element

\begin{equation}
S_0 = \df{1 - \i \beta/\alpha}{1 + \i \beta/\alpha} 
    = \df{1 - k\sqrt{P}}{1 + k\sqrt{P}} \ .                 \label{S04d}
\end{equation}

\noindent
The Hawking rate is computed as before:

\begin{equation}
\Gamma(\omega) \df{\d \omega}{2 \pi} = 
  \df{1 - |S_0|^2}{\e^{\beta_H (\omega - e a_0)} - 1} \df{\d \omega}{2 \pi} 
 \approx \df{1}{2 \pi^2} 
         \df{k (\omega - e a_0) A_{\rm h}}{\e^{\beta_H (\omega - e a_0)} - 1}
\d \omega\ ,                                                \label{Hawk4d}
\label{crate}
\end{equation}

\noindent
where we have used \equno{Area4d}.
\equno{Hawk4d} holds far 
from extremality, provided $\mu m \ll 1$.

Now we would like to argue that, for the scalars arising in four 
dimensions from 11-dimensional gravitons polarized in the 
$y_2,\ldots,y_7$ directions, this emission rate is exactly 
reproduced by a simple model analogous to the one used for
the $D=5$ black hole. This model is based on a long string 
whose left and right moving fluctuations collide to produce outgoing
charged scalars. From the M-theory point of view, this long string
pointing in the $\hat 1$ direction
is at the triple intersection of 5-branes. If the length of the 
$y_1$ direction is $L_1$, then the effective length of the 
intersection string is $\Leff=n_1 n_2 n_3 L_1$. If we assume that
there are four massless bosons and four massless fermion modes on this
long string, then the Bekenstein-Hawking entropy \equno{Area4d}
is correctly reproduced \cite{kt,kt1}. 

While the details of the dynamics of intersecting M-branes are unknown,
our calculation is largely independent of them. We just assume 
the geometrical coupling of
the long string to the gravitons polarized in the 
$y_2, \ldots, y_7$ directions:

\begin{equation}
T_{\rm D} \int \d^2 \xi \, \tf{1}{2} (\delta_{ij} + 2 \kappa_{5} h_{ij})
        \partial_\alpha X^i \partial^\alpha X^j \ .
\end{equation}

\noindent
This is identical to the coupling in \equno{StringAction}, but
with $\kappa_6$ replaced by $\kappa_5$. In fact, the entire calculation
of the emission from a long string, presented in section 2, carries over
to the $D=4$ case with minor alterations. The only changes needed
in \equno{GammaUniv} are the replacement of $\kappa_5$ by
$\kappa_4$ and of $\d^4 k/(2\pi)^4$ by $\d^3 k/(2\pi)^3$.  One then 
obtains

\begin{equation}
\Gamma (\vec{k}) \df{\d^3 k}{(2 \pi)^3} = 
  \df{\kappa_4^2 \Leff}{4} \df{k^2}{\omega} 
  \rho_L \left( \df{\omega + e}{2} \right) 
  \rho_R \left( \df{\omega - e}{2} \right) 
  \df{\d^3 k}{(2 \pi)^3}                             \label{newrate}
\end{equation}

\noindent
for the differential rate in $D=4$.
Making use of \equno{RhoLExpand}, and of \equno{temps} with $G_5$
replaced by $G_4$ (which is applicable because we again assume that
there are four species of massless bosons and fermions), 
we bring the differential rate into the form

\begin{equation}
\Gamma(\omega) \df{\d \omega}{2 \pi} = \df{1}{2 \pi^2} 
  \df{A_{\rm h} k (\omega - e)}{\e^{\beta_{\rm H} (\omega - e)} - 1}
\d \omega \ ,
\end{equation}

\noindent
which is identical to \equno{crate} derived in the context of
semi-classical gravity! While a better derivation of the emission
model based on the multiply wound string is clearly necessary,
this model incorporates important properties
of semi-classical black holes.

\section{Discussion}

In this paper we have presented new evidence in favor of a microscopic
picture behind semi-classical near-extremal black holes.  There are two
simple classes of black holes whose extremal limits preserve some
supersymmetry and are characterized by finite horizon area.  These are the
$D=5$ black hole with three $U(1)$ charges and the $D=4$ black hole with
four $U(1)$ charges.  Both cases may be represented as branes intersecting
along a string:  in the former case it is sufficient to use D-branes alone,
while in the latter one may use triply intersecting 5-branes of M-theory.
We have confirmed that microscopic models based on these brane descriptions 
predict a charged particle emission rate which agrees exactly, including
the normalization, with the semi-classical treatment of Hawking radiation.
Once the brane calculations are reduced to small fluctuations of the
intersection string, they become quite simple and almost entirely
independent of the details of the higher dimensional theory.

The formulae for charged particle emission reveal some simple
physically expected properties. If the black hole is positively charged,
then the emission of positively charged particles of mass $m$
is enhanced compared to that of negatively charged particles
of the same mass.  In fact, the emission rate for negatively charged
particles of long wavelengths contains an exponential suppression factor,
$\e^{-2 \beta_{\rm H} m}$, consistent with the semi-classical interpretation
of tunneling.

It would be interesting to calculate the
net emission rate for charge and compare it with 
the energy emission rate. Presumably, after radiating some charge
and mass, the near-extremal black hole will stabilize at some
values of charge and mass which satisfy the extremality relation.
We hope to return to this issue in a later publication.

\section*{Acknowledgements}

We are grateful to C.G.~Callan, J.~Maldacena, G. Mandal, S. Mathur,
and especially A.~Tseytlin for illuminating discussions.  The work of
I.R.~Klebanov was supported in part by DOE grant DE-FG02-91ER40671,
the NSF Presidential Young Investigator Award PHY-9157482, and the
James S.{} McDonnell Foundation grant No.{} 91-48.  S.S.~Gubser 
also thanks the Hertz Foundation for its support.




\begin{thebibliography}{99}
\frenchspacing

\bibitem{jb} J.~Bekenstein, Lett. Nuov. Cim. {\bf 4} (1972) 737;
Phys. Rev. {\bf D7} (1973) 2333; Phys. Rev. {\bf D9} (1974) 3292.

\bibitem{swh} S.W.~Hawking, Nature {\bf 248} (1974) 30;
Comm. Math. Phys. {\bf 43} (1975) 199.

\bibitem{sv} A.~Strominger and C.~Vafa, Harvard and Santa Barbara
preprint HUTP-96-A002 (January 1996), hep-th/9601029.

\bibitem{polchnotes} For a review, see
J.~Polchinski, S.~Chaudhuri, and C.V.~Johnson,
``Notes on D-Branes,'' Santa Barbara and ITP preprint NSF-ITP-96-003
(February 1996), hep-th/9602052.

\bibitem{cm} C.G.~Callan and J.M.~Maldacena, Princeton preprint
PUPT-1591 (February 1996), hep-th/9602043.

\bibitem{hs1} G.T.~Horowitz and A.~Strominger, Santa Barbara preprint
(February 1996), hep-th/9602051.

\bibitem{hk} A.~Hashimoto and I.R.~Klebanov, 
Phys. Lett. {\bf B381} (1996), hep-th/9604065.

\bibitem{dmw} A.~Dhar, G.~Mandal and S.~R.~Wadia, Tata preprint 
TIFR-TH-96/26, hep-th/9605234.

\bibitem{dm} S.R.~Das and S.D.~Mathur, MIT and Tata preprint
MIT-CTP-2546, TIFR-TH/96-19, hep-th/9606185.

\bibitem{dm1} S.R.~Das and S.D.~Mathur, MIT and Tata preprint
MIT-CTP-2508 (January 1996), hep-th/9601152.

\bibitem{ms} J.M.~Maldacena and L.~Susskind, hep-th/9604042.

\bibitem{cy} M.~Cveti\v c and D.~Youm, Phys. Rev. {\bf D53} (1996) 584,
hep-th/9507090.

\bibitem{ct} M.~Cveti\v c and A.~Tseytlin,
Phys. Lett. {\bf B366} (1996) 95, hep-th/9510097.

\bibitem{sm} J.~Maldacena and A.~Strominger, hep-th/9603060.

\bibitem{jkm} C.~Johnson, R.~Khuri and R.~Myers, hep-th/9603061.

\bibitem{at} A.~Tseytlin, hep-th/9604035.

\bibitem{kt} I.R.~Klebanov and A.~Tseytlin, hep-th/9604166.

\bibitem{gkt} J.P.~Gauntlett, D.~Kastor and J.~Traschen,
hep-th/9604179.

\bibitem{mprivate} J.M.~Maldacena
and A.~Strominger, private communication.

\bibitem{at1} A.~Tseytlin, Mod. Phys. Lett. {\bf A11} (1996)
689, hep-th/9601177.

\bibitem{kt1} I.R.~Klebanov and A.A.~Tseytlin, Princeton and Imperial 
preprint PUPT-1639, Imperial/TP/95-96/57, hep-th/9607107.

\bibitem{ct1} M.~Cveti\v c and A.~Tseytlin, hep-th/9606033.

\bibitem{hms} G.~Horowitz, J.~Maldacena and
A.~Strominger, hep-th/9603109.

\bibitem{unruh} W.G.~Unruh, Phys. Rev. {\bf D14} (1976) 3251.

\bibitem{gibbons} G.W.~Gibbons, Comm. Math. Phys. {\bf 44} (1975) 245. 


\end{thebibliography}
\end{document}